\documentclass[10pt]{article} 
\usepackage{simpleConference}
\usepackage{times}
\usepackage{graphicx}
\usepackage{amssymb}
\usepackage{float}
\usepackage{multicol}

\begin{document}

\title{\textbf{Propagation of Cyclonic Vortices and Intense Rainfall over Indian Peninsula: Case Studies from Northeast Monsoon Season}}

\author{Jayesh Phadtare \\
\small{Centre for Atmospheric and Oceanic Sciences, 
              Indian Institute of Science, Bengaluru,
              India - 560012}\\
 \date{}
\\
\small{e-mail: phadtare@iisc.ac.in} \\
}

\maketitle

\begin{multicols}{2}

\begin{abstract}
Four case studies of cyclonic vortices over the Indian peninsular region during northeast monsoon season are presented. It was observed that the propagation of vortices over the peninsular orography was controlled by the vortex Froude number, $Fr_{v} = V_{max}/(NH)$, where, $V_{max}$ is the maximum tangential velocity of winds in the vortex, $H$ is the height of the orography, and $N$ is the Brunt-V\"{a}is\"{a}l\"{a} frequency of atmosphere along the orography. Strong vortices with $Fr_{v}>1$ overcame the orographic barrier and crossed the peninsula.  While, those with $Fr_{v}<1$ were blocked by the orography. The blocked vortices remained stationary along the east coast  for two to three days. In these cases, clouds formed continuously along the coast for three consecutive days. Therefore, there was intense rainfall over the coast and little rainfall over orography during this time. In unblocked vortices, the rainfall intensity showed a systematic shift from  coastal to the higher orographic region within a day or two. However, significant weakening of the vortices, as well as convection, was seen as the vortex arrived over the higher plateau. These case studies will be helpful for forecasting  heavy rainfall events over the Indian peninsula during northeast monsoon season.
\end{abstract}

\section{Introduction}
\label{intro}
\begin{figure}[H]
  \includegraphics[height= 3.0in, width=7in]{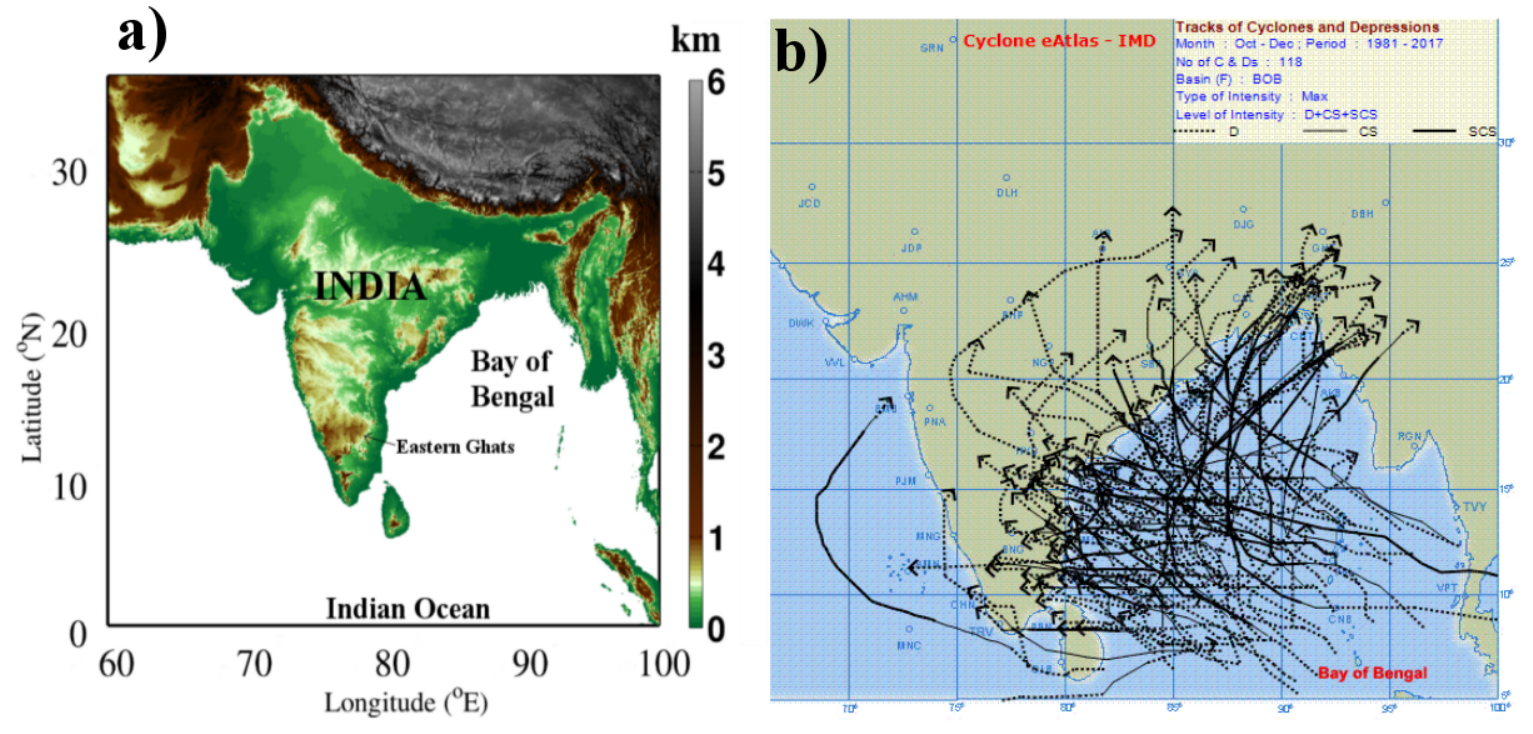}
\caption{a) Elevation from the ETOPO5 (5-minute gridded elevation) data b) Tracks of depressions and tropical cyclones formed over BoB during October-December 1981-2017 (plotted from IMD's cyclone eAtlas web-portal : http://www.rmcchennaieatlas.tn.nic.in/).}
\label{fig:tracks} 
\end{figure}

Tropical disturbances over the Indian ocean often develop in the form of low pressure systems (LPSs); winds flow around LPSs in cyclonic fashion obeying the gradient wind balance in which the inward pressure gradient force is balanced by the Coriolis and centrifugal force \cite{holton2012chp3}.  These cyclonic vortices, in the absence of strong background flow, move westward-northwestward due to the planetary $\beta$-effect \cite{chan1987,fiorino1989,wang2004,chan2005}.
The vortices over Indian region are classified according to their strength by the India Meteorological  Department (IMD) in the following categories:- low pressure area (LPA), depression (Dp), deep depression (DDp), and  four categories of tropical cyclone (TC) thereafter (table \ref{tab:LPSs}). During northeast monsoon (October-December), these cyclonic systems predominantly form over the southern Bay of Bengal (BoB) and the eastern equatorial Indian ocean. These systems then move westwards towards the Indian peninsula; some of them curve back north-eastwards and make landfall over the Bangladesh-Myanmar region (figure \ref{fig:tracks}b). 
During this time of the year, a 30-50 m deep warm layer of fresh water from river run-offs and summer monsoon rainfall sits over BoB and fuels the growth of these vortices \cite{sengupta2008}. They get better organized with the time they spend over the warm ocean, if not weakened by the vertical wind shear \cite{frank2001}. A vortex of TC strength can cause widespread damage on landfall due to heavy rainfall, strong winds and the storm surge. Slow moving or quasi-stationary systems, even with weak organization, can deposit large amount of rainfall at a location and cause severe local flooding \cite{schumacher2009quasi,phadtare2018}

As far as the prediction of TC is concerned, prediction of cyclone track with numerical weather models has improved over the last couple of decades globally  \cite{leroux2018}, including Indian region \cite{vaidya2004,mohapatra2013,osuri2015,mohanty2016,kumkar2018,routray2018}. On the other hand, the long duration localized heavy rainfall episodes over the Indian east coast due to quasi-stationary vortices has not received much attention. In some cases, loss of life and property due to quasi-stationary precipitating system can be far sever. A quasi-stationary precipitating system on 1 December 2015 deposited 490 mm of rainfall over Chennai within 24 hrs (\cite{phadtare2018}). This resulted in disastrous flooding in the city \cite{chennai2016}. Armed and disaster relief forces were deployed for relief and rescue operations. At least 250 people died in the calamity. Prior to this particular catastrophic event, two extreme rainfall events occurred in Tamil Nadu state in November 2015 \cite{chakraborty2016}).  All three events were caused by the cyclonic systems of weak to moderate strength. No TC formed over BoB in this season. This fact highlights the need for a better understanding of these systems and the conditions in which they become quasi-stationary. 

Over Indian peninsula, the Eastern Ghats (EG) mountain range (average height $\simeq$ 750 m) lies parallel to the east coast, 200 km inland from it (figure \ref{fig:tracks}a).  The upstream orographic blocking of easterly winds by the EG mountains and the subsequent stagnation of precipitating clouds was the reason for the extreme rainfall event of 1 December 2015 \cite{phadtare2018}. The cyclonic system remained stationary along the coast for about 48 hrs. Thereafter, it traveled a southward path along the coast. This behaviour of westward moving cyclonic system when blocked by the orography is peculiar; it is caused by the topographic $\beta$-effect  \cite{carnevale1991,zehnder1993,kuo2001}. Such blocking of cyclonic vortices is also observed along the Central Mountain Range of Taiwan \cite{lin2005} and the Sierra Madre mountains in Mexico \cite{zehnder1993}.  

The fate of a vortex approaching an orography is determined by the vortex Froude number,

\begin{equation}\label{eqn:Fr}
Fr_{v} = \frac{V_{max}}{NH}       
\end{equation}
where, $V_{max}$ is the maximum tangential velocity of winds in the vortex, $H$ is the height of the orography, and $N$ is the Brunt-V\"{a}is\"{a}l\"{a} frequency of atmosphere along the orography \cite{lin2005}. $N$ is a measure of  stratification of atmosphere and is given by \cite{emanuel1994},

\begin{equation}\label{eqn:N}
N = [(\frac{g}{\theta_{vo}})(\frac{\partial{\theta_{v}}}{\partial{z}})]^{1/2}
\end{equation}
where, $g$ is the gravity, $\theta_{v}$ is the virtual potential temperature and $\frac{\partial{\theta_{v}}}{\partial{z}}$ is the vertical gradient of $\theta_{v}$.  $Fr_{v}$ is analogous to the conventional Froude number \cite{smith1979,pierrehumbert1985,lin1996} in which the cross-mountain wind speed term is replaced by the $V_{max}$. $Fr_{v}$ essentially is the ratio of prevailing wind kinetic energy within the vortex and the minimum kinetic energy requirement for the winds to climb the orographic barrier. Provided that the orography is sufficiently long (in a direction perpendicular to the vortex propagation), vortices with $Fr_{v} <1$ get blocked and those with $Fr_{v} > 1$ cross the orographic barrier.

This paper presents case studies of interaction of westward moving cyclonic systems, of weak to moderate strength during 8-10 November 2015 (hereafter, Case-1), 14-16 November 2015 (hereafter, Case-2) and 31 October - 2 November 2017 (hereafter, Case-3) with the EG orography. 3-day rainfall accumulation along the coast in these cases was $\geq$ 300 mm at some places over the coast.  A similar analysis for TC Vardah (hereafter, Case-4) which made landfall near Chennai on 12 December 2016 is also presented. Organization of convection and distribution of rainfall within these systems is also studied. A comparison of spatio-temporal distribution of rainfall over the peninsular region due to these systems emphasizes  the importance of understanding the stalling dynamics of weak systems along the coast. 

Intense rainfall events over the west coast of India during summer monsoon and the associated weather systems  are well documented (\cite{francis2006}). Similar studies on the intense rainfall events over the east coast of India during northeast monsoon are scarce. Such an exercise will be helpful to the forecasters for forecasting heavy rainfall events over Indian peninsula due to  winter-time cyclonic systems. In addition, better understanding of the vortex-orography interaction is necessary for the comprehension and improvement of the forecast of numerical weather models. 

Section \ref{sec:data} of this paper provides the description of data used in this study. Section \ref{sec:blocking} presents the stalling dynamics of vortices; here, the organization of convection and rainfall distribution in four case studies is also presented. Section \ref{sec:discussion} contains a discussion on the results and section \ref{sec:conclusion} concludes the study.

\section{Data}\label{sec:data}
ERA-interim dataset \cite{dee2011} with $0.75^{\circ}$ and 6 hourly resolution provided by the European Centre for Medium-Range Weather Forecasts (ECMWF) is used to analyze the synoptic dynamics of vortices in four cases. Radiosonde sounding data from Chennai is used to measure the wind speed and stratification ($N$) of atmosphere along the coast. Infrared (IR $\sim$ 10.8 $\mu m$) images from the INSAT-3D satellite are used to study the organization of deep convection within the vortices.
This geostationary satellite is located over $82^{\circ}$E. Its IR images have 4 km spatial and 30 minute temporal resolution. 
These images are archived by the Meteorological and Oceanographic Satellite Data Archival Centre, Satellite Applications Centre, Indian Space Research Organization (ISRO). Integrated Multi-satellitE Retrievals for GPM (IMERG) data product provided by the National Aeronautics and Space Administration (NASA) is used for rainfall.  IMERG has 10 km spatial and 30 minute temporal resolution. The tracks of depressions and TCs shown in figure \ref{fig:tracks} are taken from IMD’s cyclone eAtlas portal.

\section{Case Studies}\label{sec:blocking}
\subsection{Dynamics}
\begin{figure}[H]
  \includegraphics[scale=1]{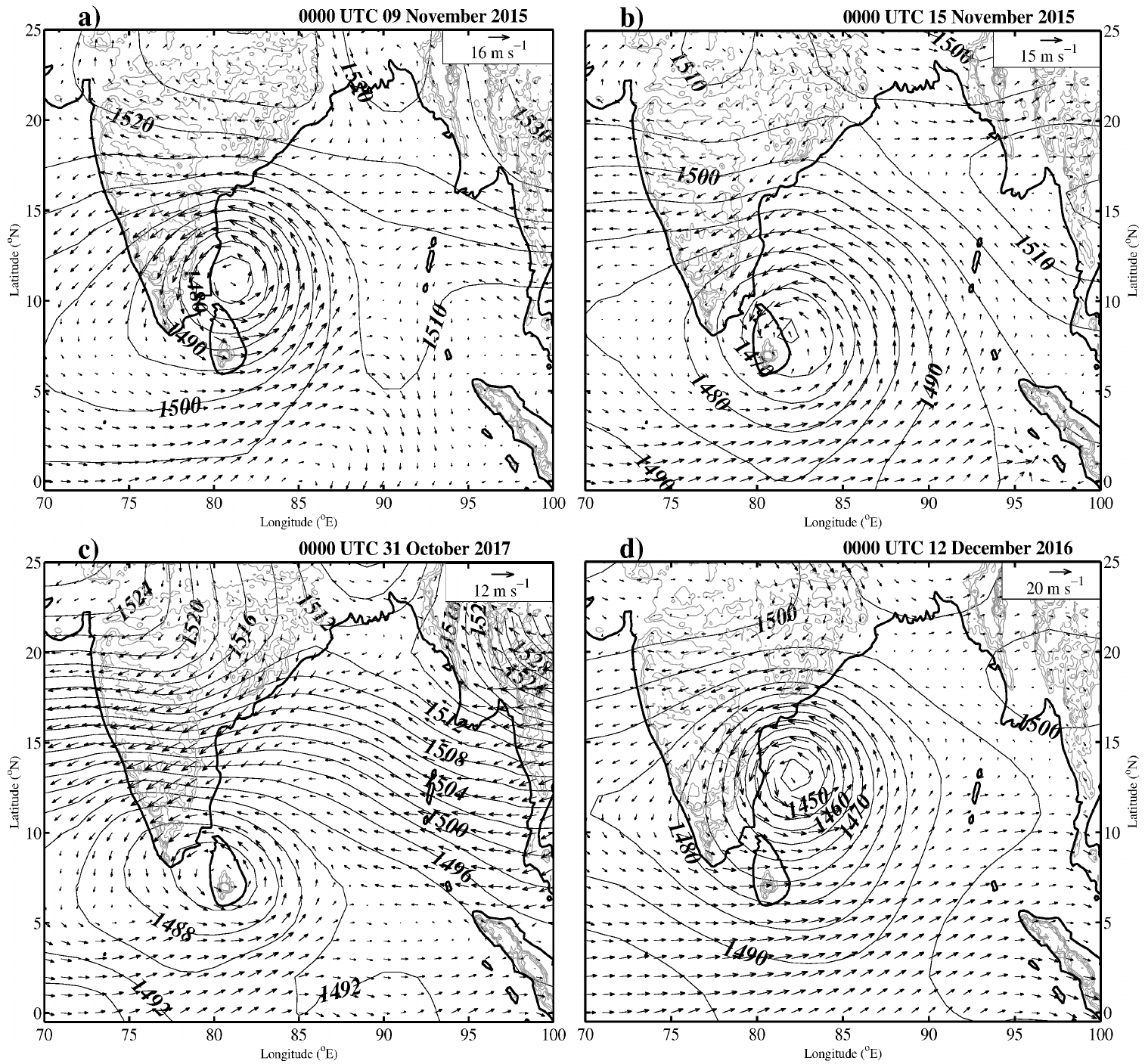}
\caption{Winds and geopotential contours at 850 hPa:  a) Case-1: 09 November 2015  b) Case-2: 15 November 2015 c) Case 3: 31 October 2017 
d) Case-4: 12 December 2016, all at 0000 UTC.}
\label{fig:initial} 
\end{figure}

\begin{figure}[H]
  \includegraphics[scale=1]{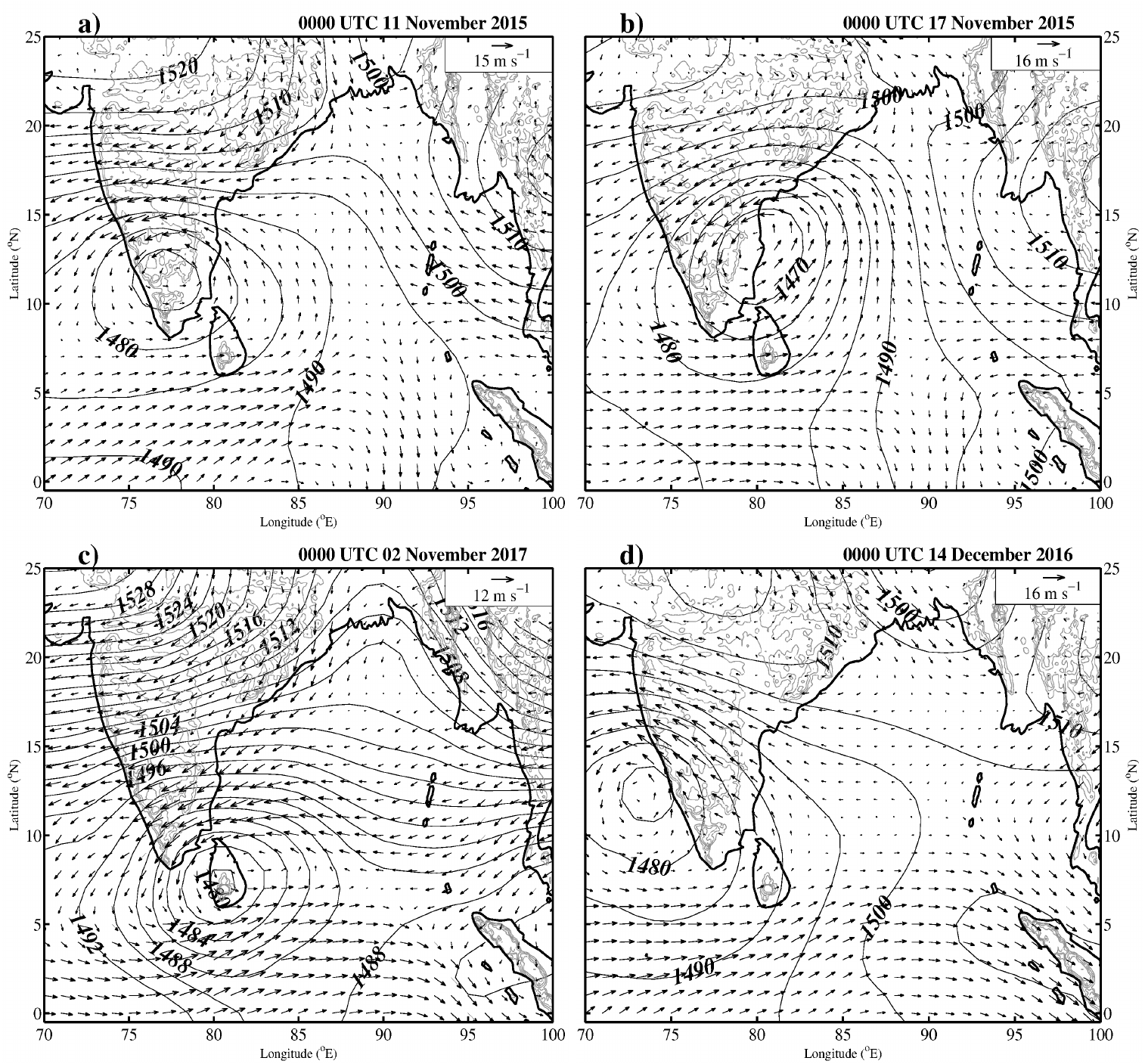}
   \caption{Same as figure \ref{fig:initial} : a) Case-1: 11 November 2015  b) Case-2: 17 November 2015 c) Case 3: 2 November 2017 d) Case-4: 14 December 2016, all at 0000 UTC.}
\label{fig:final} 
\end{figure}

\begin{figure}[H]
  \includegraphics[scale=1.2]{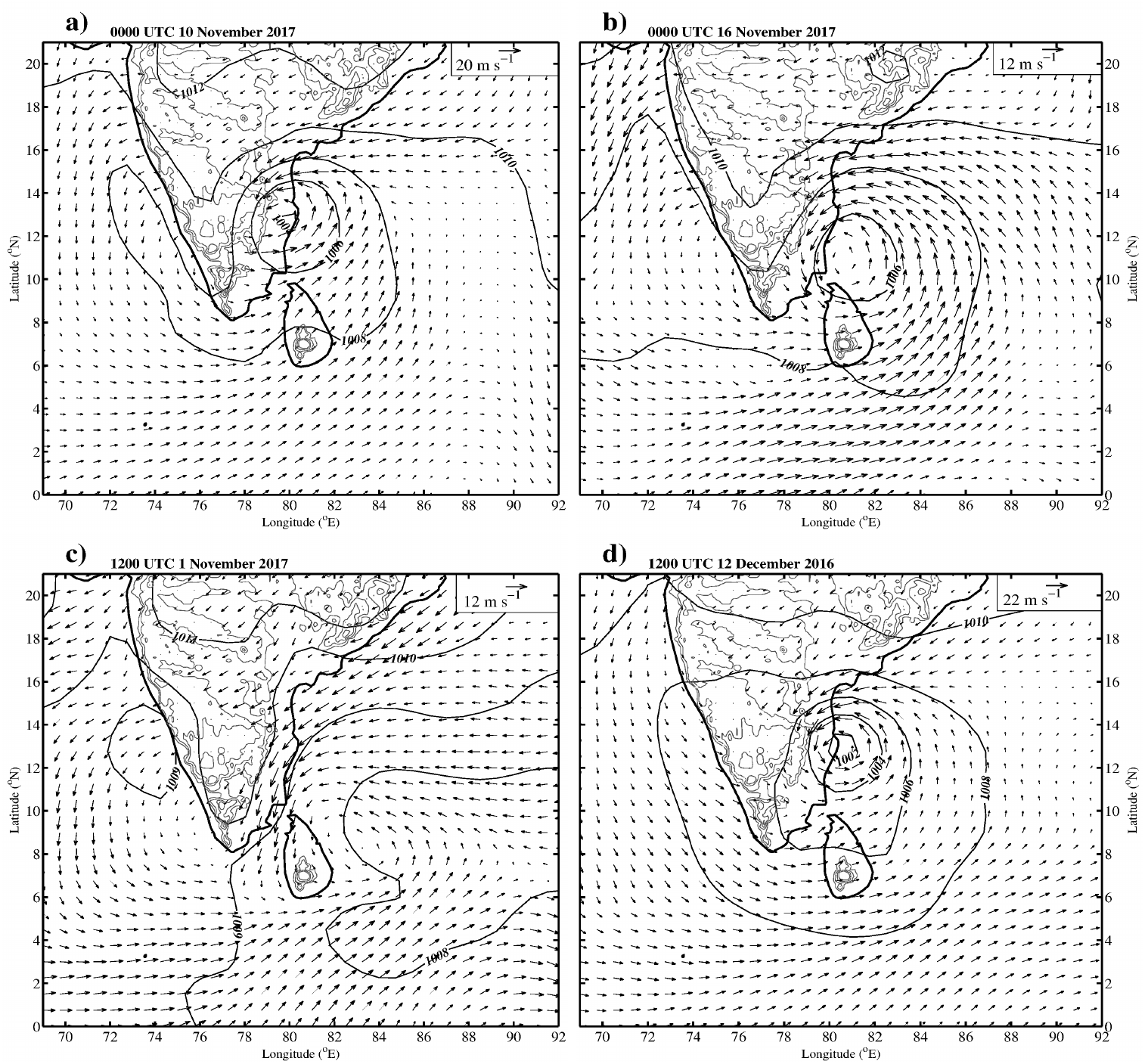}
   \caption{Mean sea level pressure contours and 950 hPa winds: a) Case-1: 0000 UTC 10 November 2015  b) Case-2: 0000 UTC 16 November 2015 c) Case 3:  1200 UTC 01 November 2017 d) Case-4: 1200 UTC 12 December 2016.}
\label{fig:950hPa} 
\end{figure}

Figure \ref{fig:initial} shows the initial landfall of vortices in four cases. Geopotential contours and winds at 850 hPa are shown in the figure. The strengths of the vortices can be inferred from the wind speeds or from the closeness in the geopotential contours. The vortex in case-1 falls in the Dp category;  the vortices in case-2, and 4 fall in LPA, and TC categories, respectively (table \ref{tab:Fr}).  The vortex in case-3 was weaker than the LPA category. It did not show any closed isobar at the surface (figure \ref{fig:950hPa}). Figure \ref{fig:final} shows  locations of these vortices after two days. Vortices in case-1 and 4, have crossed the coast by this time. In fact, TC Vardah in case-4 has crossed the peninsula and moved over Arabian sea.  Dp in case-1 has moved over the peninsular plateau. Note that, these two vortices have become weaker while crossing the peninsula. Vortices in case-2 and 3 are still along the coast. Figure \ref{fig:950hPa} shows the mean sea level pressure contours and 950 hPa winds. The height of 950 hPa surface is roughly 500 m; easterly winds at this level impinge on the EG orographic barrier. In case-1 and 4 winds are strong with speeds around 20 $m$ $s^{-1}$. Vortices in these two cases have moved inland. Whereas, the vortices in case-2 and 3 have weak winds with speeds around 12 $m$ $s^{-1}$. These two vortices remained along the coast. In case-3, the vortex circulation features are not so well developed at the 950 hPa unlike the 850 hPa level. Therefore, blocking of easterly winds and its southward deflection is more evident in this case (figure \ref{fig:950hPa}c).

Table \ref{tab:Fr} shows that the $Fr_{v}$ values (calculated from the Chennai radiosonde data) in case-1 and 4 are greater than 1. The $Fr_{v}$ values suggest that these vortices will cross the EG orography. Note that the TC Vardah crossed the coast relatively faster. The radiosondes -- which are released every 12 hours and give a point measurement -- could not measure the peak wind speeds of TC Vardah. A very severe cyclonic storm, like Vardah, can have peak wind speeds around 40-50 $m$ $s^{-1}$. This suggests that the $Fr_{v}$ values of land-falling TC of this category can be around 3-3.5. In fact, in all cases the $Fr_{v}$ values calculated from the soundings must be an underestimation of real $Fr_{v}$ values, as the vortices are already under the influence of the orography.
The upstream pressure perturbations due to the orography decelerates the flow ahead of it \cite{markowski2011}. The true $Fr_{v}$ values should be calculated when the vortex is away from these orographic effects. The distance up to which an orography can have decelerating influence on the winds is given by the orographic deformation radius \cite{pierrehumbert1985}, 
\begin{equation}\label{eqn:rossby}
D = \frac{NH}{f}
\end{equation}
where, $f$ is the Coriolis parameter. For EG (considering $f = 0.3 \times 10^{-4}$ $s^{-1}$, $H$ = 750 $m$, and $N$ = 0.01 $s^{-1}$), this distance is about 250 $km$.  The southerly barrier jet along the coast due to the  orographic deflection of 950 hPa easterly winds show similar width in figure \ref{fig:950hPa}c. Therefore, the coastal atmosphere at 200 $km$ away from the orography is under its influence. Due to the lack of observations over the open ocean, the soundings from a coastal station are used here for calculating $Fr_{v}$. Nevertheless, the calculations done in  table \ref{tab:Fr} suggest that vortices with $Fr_{v}$ $>$ 1 overcame the orographic barrier and those with $Fr_{v}$ $< $ 1 were blocked.

\subsection{Convection}
\begin{figure}[H]
  \includegraphics[scale=1]{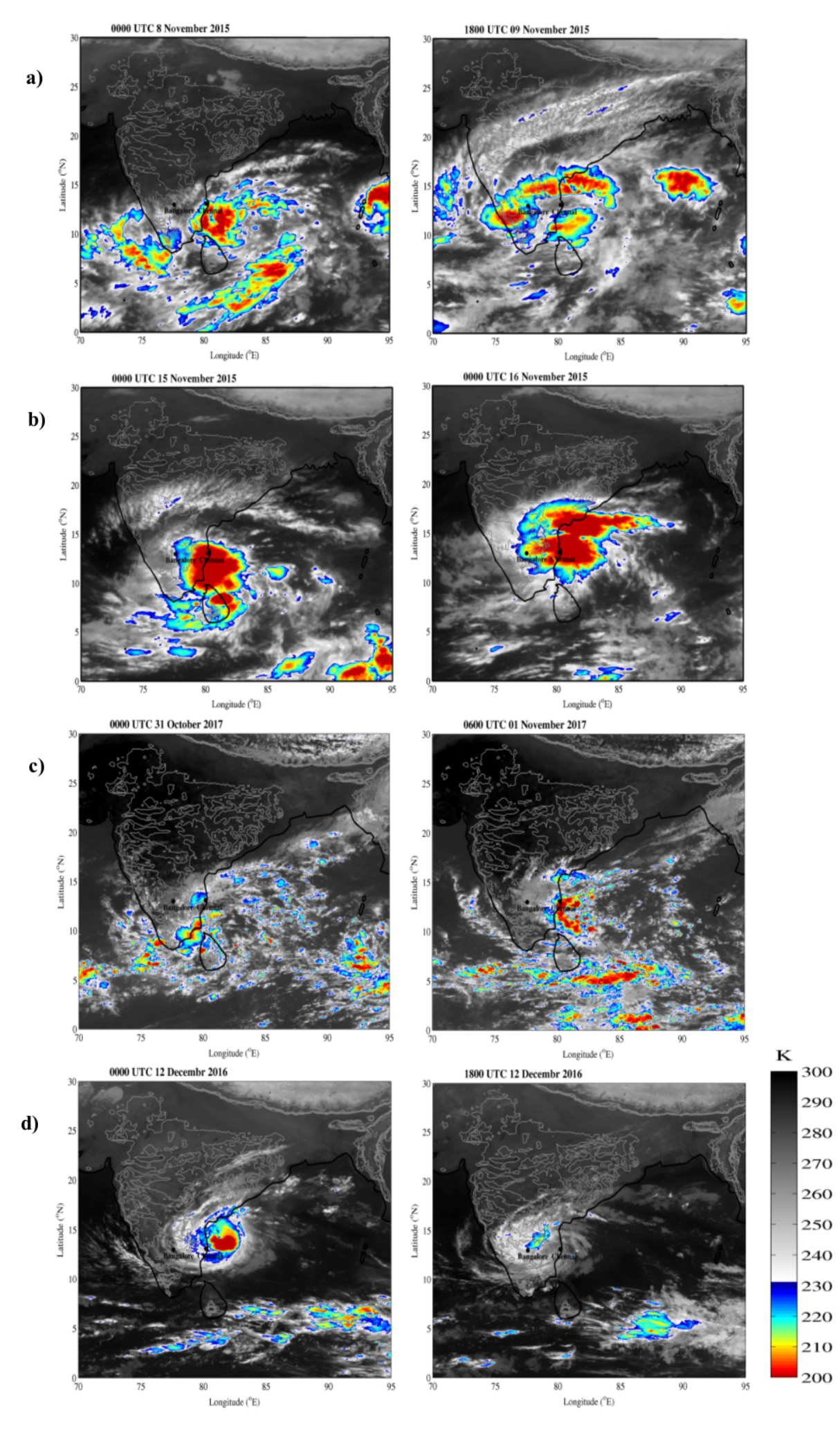}
   \caption{INSAT-3D IR brightness temperature: a) Case-1 b) Case-2 c) Case-3 d) Case-4}
\label{fig:IR} 
\end{figure}

Figure \ref{fig:IR} shows the organization and propagation of deep clouds in these four cases. In case-1, clouds cross the EG mountains (figure \ref{fig:IR}a). Clouds are organized in a more or less circular fashion along the coast on 8 November. On next day, after crossing the mountains, the cloud organization has broken down to some extent. However, the deep convection still covers a large area of peninsula. In case-2, the clouds are highly organized and anchored along the coast on 15 November (figure \ref{fig:IR}b). They remained anchored and organized over the coast even on the next day. Similar stagnation of clouds over the coast was observed on 1 December 2015 when the Froude number of the cross-mountain winds was less than 1 \cite{phadtare2018} . The study proposed a mechanism in which the upstream orographic blocking and uplifting of winds result in continuous building-up of clouds along the coast. Similar building-up of clouds along the coast must have taken place in case-2. In case-3 ($Fr_{v} < 1$) also the clouds are anchored over the coast but they cover lesser area and have weak organization (figure \ref{fig:IR}c). In case-4, the cloud organization during the landfall is very compact; the spiral rain-bands around the core have weakened, the eyewall at the center also has lost its structure. The cloud blob crosses the EG mountains within a day. However, the deep convection looses its strength and is very feeble over the plateau. 

In summary, the clouds in the blocked vortices remained anchored over the coast for more than a day. Whereas, the clouds in unblocked vortices moved over the orography, but lost its organization at the same time.
\subsection{Rainfall}
The different nature of organization and propagation of clouds result in different spatio-temporal distribution of rainfall over the region in these four cases. Note that the total accumulation of rainfall at a location depends on the intensity and the duration of rainfall episode \cite{doswell1996}. The intensity depends on the factors like cloud-microphysics or rainfall efficiency; this study does not intent to cover this aspect. The emphasis of this study is on the `duration' aspect of rainfall which mainly depends on the propagation of storm.

\begin{figure}[H]
  \includegraphics[scale=1]{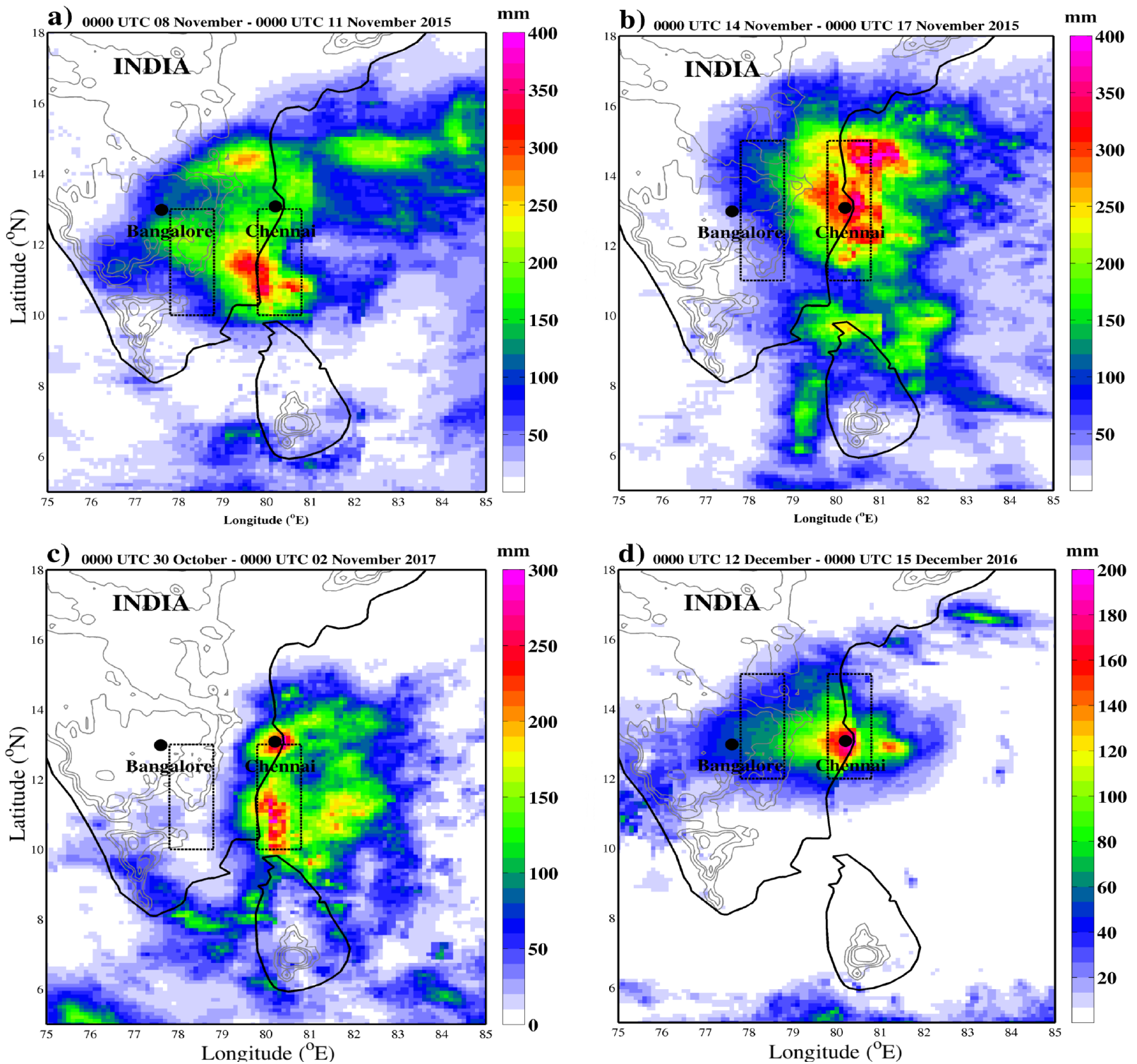}
   \caption{IMERG 3-day rainfall accumulation: a) Case-1  b) Case-2 c) Case 3 d) Case-4. The boxes show the areas over which rainfall is averaged to produce time-series in figure \ref{fig:timeseries}. }
\label{fig:gpm} 
\end{figure}

Figure \ref{fig:gpm} shows 3-day accumulation of rainfall over peninsula from the GPM satellite for the four cases. In case-1, the maximum value of rainfall accumulation over the coast is about 300 mm and over plateau its about 200 mm. Whereas, in case-2 -- in which there was a stagnation of clouds and the vortex along the coast -- the maximum value of rainfall accumulation over the coast is about 400 mm and over plateau its about 100 mm. Similarly in case-3 (a blocked system), rainfall is anchored along the coast and over plateau there is scarce rainfall. In case-4 (an unblocked system), we see significant rainfall accumulation over the plateau along with the coast. 

\begin{figure}[H]
  \includegraphics[scale=1]{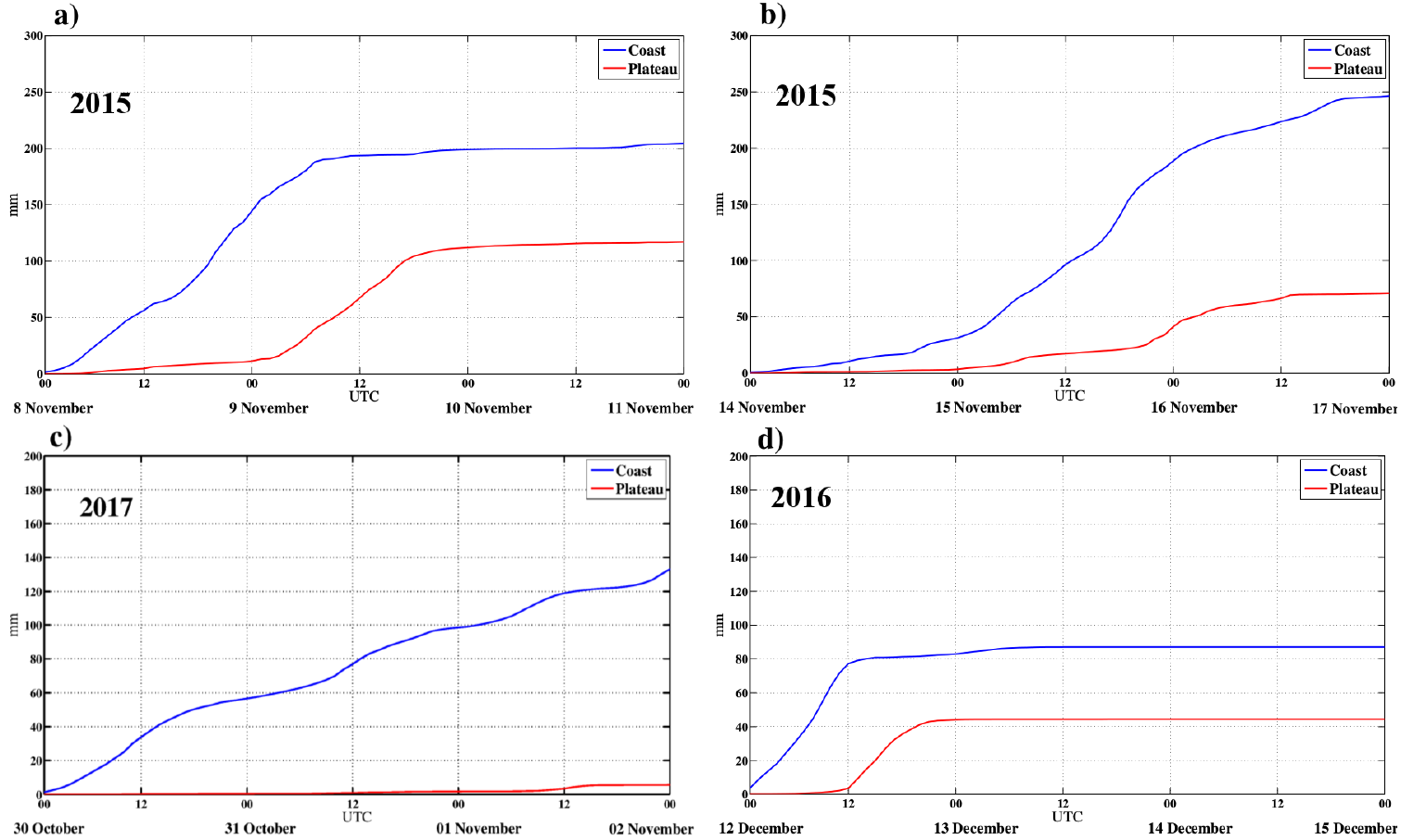}
   \caption{ Time-series of mean IMERG rainfall accumulation over coast and plateau: a) Case-1  b) Case-2 c) Case-3 d) Case-4. Rainfall is averaged over the boxes shown in figure \ref{fig:gpm} }
\label{fig:timeseries} 
\end{figure}

A 3-day time-series of rainfall accumulation over coast and plateau during the lifetime of these systems, brings out the difference in rainfall accumulations for blocked and unblocked systems more clearly. The boxes (one each over coast and plateau) for the four cases over which the rainfall is averaged are shown in figure \ref{fig:gpm}. The time-series of mean rainfall over these boxes are shown in figure \ref{fig:timeseries}.  In case-1, it rains for about 30 hrs over the coast from 8 November; and thereafter there is hardly any rainfall (figure \ref{fig:timeseries}a). Over plateau, the rainfall picks-up on 9 November, about a day later than the coast. Here, it rains for about 18 hours and then it stops thereafter. The two time-series in case-1 suggest a systematic movement of precipitating system crossing the coast and  plateau on subsequent days. In case-2, the vortex and convection were stationary over the coast. Therefore, the rainfall time-series over the coast shows continuous occurrences of rainfall for three consecutive days (figure \ref{fig:timeseries}b). Whereas, over the plateau the rainfall accumulation is moderate and does not show any steep rise at any point. Similar conclusions can be drawn for the two timeseries in case-3 (figure \ref{fig:timeseries}c). In case-4, the time-series at two locations are similar to those in case-1, but the duration of rainfall is much shorter and the rainfall over plateau picks-up just 12 hours later than that over coast.  This is due to the fact that the TC in case-4 moved across the peninsula more swiftly than the Dp in case-1.

In summary, in the cases of blocked vortices (case-2 and 3), convection was stationary over the coast for two-three days. This lead to continuous rainfall accumulation over the coastal region for three days and scarce rainfall over the plateau. Strong vortices (case-3 and 4) traveled across the peninsula overcoming the orographic barrier. In these cases, rainfall lasted for a day over the coast and plateau; coastal regions received more rainfall than the plateau. 

\section{Discussion} \label{sec:discussion}
It was observed that the vortices with $Fr_{v} > 1$ crossed the peninsula and those with $Fr_{v} <1$ were blocked by the EG orography. The vortices that moved over the orography were weakened while doing so. The convection also lost its organization at the same time. This can happen due to the lack of moisture over the plateau and friction due to the uneven terrain surface. Another reason can be the compression of vorticity column as the vortex moves over the plateau. Potential vorticity (PV) of a barotropic vortex is conserved \cite{holton2012chp4}:
\begin{equation}\label{eqn:pv}
\frac{D}{Dt}\left(\frac{f + \zeta}{h} \right) = 0
\end{equation}
where, $\zeta$ is the vertical component of relative vorticity and $h$ is the height of the vortex column.
The term within the brackets on the left hand side of equation \ref{eqn:pv} is the PV of the vortex.
Thus, as the $h$  is reduced when the vortex moves over the plateau, there should be a corresponding decrease in $\zeta$, so that the PV is conserved. However, the dominant mechanism by which the vortex is weakened can not be determined on purely observational account. Modeling experiments can shed some light on this aspect.
 
Interestingly, in case of blocked vortices, the orographic blocking of winds at the bottom layer of atmosphere (depth of about 750 m) 
resulted in the entire column of the vortex getting blocked. This behaviour is similar to the 2-D  `Taylor columns' or `Produman pillars' in which the entire fluid column behaves as a rigid entity \cite{pedlosky2013}. 
However, approximation of tropical vortices by Taylor columns is arguable, given the strong convective activity and hence, the vertical motion associated with the former. 
Latent heating due to the clouds leads to the generation of vorticity within Dp \cite{rao1970,sanders1984,chen2005}.
Westward propagation of Dp during the Indian summer monsoon season is linked to this in-situ vorticity generation by these studies. Within the blocked vortices studied in this paper, clouds were stagnated and repeatedly formed along the coast. Thus, the vorticity generation can happen at the same place over a time and the westward movement of vortex can get seized. Identification of the exact mechanism by which the motion of entire vortex is halted -- whether by Taylor column effect or by the convective heating -- is also important to determine the role mesoscale convection within the depressions.

In case-3, it was seen that the TC Vardah had a compact cloud organization during the landfall. Whereas, vortices in case-1 and case-2 had widespread deep cloud cover (figure \ref{fig:IR}). Thus, the rainfall in these two cases was more widespread compared to TC Vardah (figure \ref{fig:gpm}). In addition, these weaker vortices were slow moving (case-2 was practically stationary along the coast). TC Vardah moved across the Indian peninsula relatively quickly. Hence, the values of 3-day rainfall accumulation over the region was more compared to TC Vardah. This demands a more careful attention of researchers and forecasters towards the weaker cyclonic vortices and the cloud organization within it. A formal methodology to forecast the stalling of such vortices along the coast should be scrupulously designed for the forecasters. 

\section{Conclusion} \label{sec:conclusion}
Propagation of four cyclonic vortices over the peninsular India during northeast monsoon season was presented. Effect of EG orography on their propagation, the cloud organization, and rainfall accumulation was studied. Following are the main conclusions drawn from this study:
\begin{enumerate}
\item The propagation of vortices over the peninsula was governed by their $Fr_{v}$ values. Vortices with $Fr_{v}>1$ 
($Fr_{v}<1$) crossed the EG orography (were blocked). TC Vardah with the greatest $Fr_{v}$ in case-4 was the fastest to cross the orography.
\item Deep convection within TC Vardah during landfall was compact and covered lesser area compared to the widespread convection within the weaker depression and LPA in case-1 and case-2, respectively. 
\item Clouds in the blocked vortices repeatedly formed along the coast, inundating the region. Rainfall over the plateau in these cases was scarce to moderate. Whereas, in the unblocked vortices deep clouds moved over the plateau and it received significant rainfall.
\item In blocked vortices, the rainfall time-series showed a continuous accumulation of rainfall over the coastal region for three consecutive days. In unblocked vortices, the rainfall time-series showed a systematic shift of rainfall from coastal to plateau region within a time-span dependent on the propagation speed of vortices.
\end{enumerate}

\bibliographystyle{abbrv}
\bibliography{my_references}
\end{multicols}

\section*{Acknowledgements}
\noindent
The author would like to thank the European Centre for Medium-Range Weather Forecasts (ECMWF) for the ERA-interim dataset; Department of Atmospheric Science, University of Wyoming for the sounding data; Meteorological and Oceanographic Satellite Data Archival Centre, Satellite Applications Centre, Indian Space Research Organization (ISRO) for providing INSAT-3D data; the National Aeronautics and Space Administration (NASA) for the IMERG rainfall data; and finally the Regional Meteorological Centre-Chennai, India Meteorological Department (IMD) for the depression and tropical cyclone tracks.
 
\begin{table}[h]
\caption{Classification chart for cyclonic systems by IMD. Isobars are plotted at the surface with 2 hPa intervals. 1 Kts $\simeq$ 0.5 $m$ $s^{-1}$}
\label{tab:LPSs}       
\begin{tabular}{lll}
\hline\noalign{\smallskip}
 System & No. of closed isobars & Wind speed \\
\noalign{\smallskip}\hline\noalign{\smallskip}
LPA & 1 &  $<$17 Kts \\
 Dp & 2-3 & 17-27 Kts \\
 DDp & 2-3 & 28-33 Kts\\
 TC & $>$ 4 & $>$ 33 Kts\\
\noalign{\smallskip}\hline
\end{tabular}
\end{table}

\begin{table}[h]
\caption{Observations for the four cases. The radiosonde soundings data from Chennai at 0000 UTC on the mentioned dates are used to calculate wind speed and $N$ values. Mean values of wind speed and $N$ within 1 km layer from the surface are shown and used to calculate $Fr_{v}$ values from equation \ref{eqn:Fr}.  $H$ is taken to be 750 m. Vertical profiles of  $N$ are shown in the figure \ref{fig:supp} }
\label{tab:Fr}       
\begin{tabular}{lllllll}
\hline\noalign{\smallskip}
Case & Category & Date  & Wind speed ($m$ $s^{-1}$) & $N (s^{-1})$ & $Fr_{v}$ & Blocking\\
\noalign{\smallskip}\hline\noalign{\smallskip}
 1 & Dp    & 09 November 2015  & 12   & 0.013 & 1.2 & Unblocked\\
 2 & LPA & 15 November 2015  & 7.5   & 0.013  & 0.77 & Blocked\\
 3 & -- & 01 November 2017  & 7   & 0.015  & 0.6 & Blocked\\
 4 & TC   &  12 December 2016  & 17 & 0.015  & 1.5 & Unblocked\\
\noalign{\smallskip}\hline
\end{tabular}
\end{table}
\clearpage

\begin{figure}[H]
  \includegraphics[scale=1]{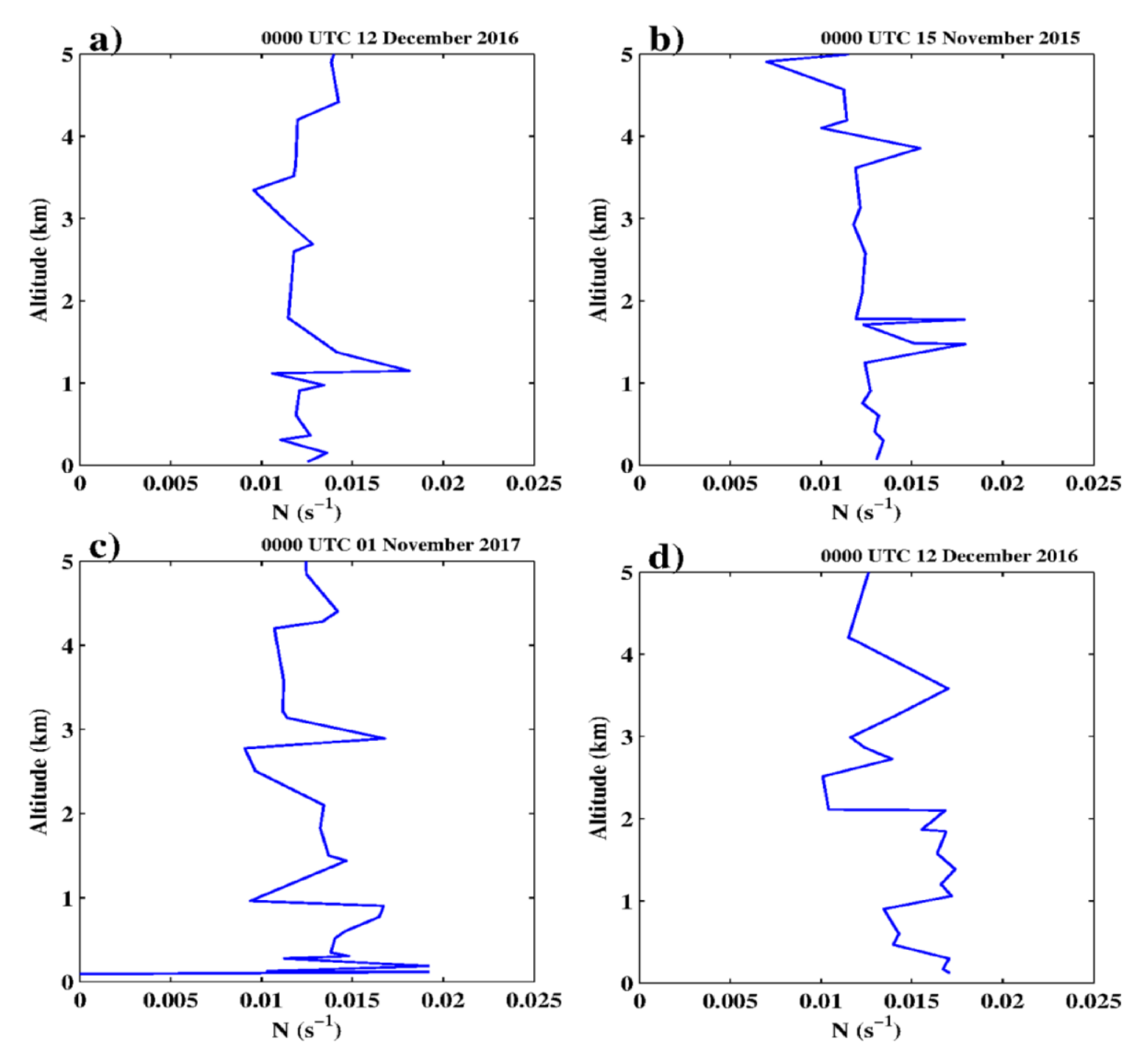}
   \caption{ Vertical profiles of Brunt-V\"{a}is\"{a}l\"{a} frequency ($N $) from the radiosonde soundings over
Chennai. a) Case-1, b) Case-2, c) Case-3, d) Case-4}
\label{fig:supp} 
\end{figure}

\end{document}